\documentclass[twocolumn,prd,nofootinbib,showpacs,preprintnumbers]{revtex4}
\usepackage{epsfig}
\usepackage{color}
\usepackage{amssymb}

\newcommand{\be}{\begin{equation}}
\newcommand{\ee}{\end{equation}}
\newcommand{\cF}{{\cal F}}

\begin{document}
\title{Grammage of cosmic rays around Galactic supernova remnants}
\author{Marta D'Angelo$\,{}^{1}$, Pasquale Blasi$\,{}^{2,1}$ and Elena Amato$\,{}^{2}$}
\affiliation{$^{1}$Gran Sasso Science Institute (INFN), Viale F. Crispi 6, 60100 L'Aquila, Italy}
\affiliation{$^{2}$INAF-Osservatorio Astrofisico di Arcetri, Largo E. Fermi, 5 50125 Firenze, Italy}

\date{\today}

\begin{abstract}
The residence time of cosmic rays (CRs) in the Galaxy is usually inferred from the measurement of the ratio of secondary-to-primary nuclei, as for instance the boron (B)/carbon (C) ratio, which provides an estimate of the amount of matter traversed by CRs during their propagation, the so called CR grammage. However, after being released by their parent sources, for instance supernova remnants (SNRs), CRs must cross the disc of the Galaxy, before entering the much lower density halo, in which they are believed to spend most of the time before eventually escaping the Galaxy. In the near-source region, the CR propagation is shown to be dominated by the non-linear self-generation of waves. Here we show that due to this effect, the time that CRs with energies up to $\sim$ 10 TeV spend within a distance $L_{c}\sim 100$ pc from the sources is much larger than naive estimates would suggest. Depending on the level of ionisation of the medium surrounding the source, the grammage accumulated in the source vicinity may be a non-negligible fraction of the total grammage traversed throughout the whole Galaxy. Moreover, there is an irreducible grammage that CRs traverse while trapped downstream of the shock that accelerated them, though this contribution is rather uncertain. We conclude that some caution should be used in inferring parameters of Galactic CR propagation from measurements of the B/C ratio.

\end{abstract}

\maketitle
%%%%%%%%%%%%%%%%%%%%%%%%%%%%%%%%%%%%%%%%%%%%%%%%%%%%%%%%%%%%%%%%%%%%%%
{\it Introduction} -- The simplest version of the standard picture of the origin and propagation of Galactic CRs is based on two pillars (see \cite{blasirev,amatorev} for recent reviews): 1) CRs are accelerated, with $\sim 10\%$ efficiency, in supernova remnants (SNRs), located in the gas disc of the galaxy, within a region of half-width $h\sim 150$ pc, where the density is $n_{d}\simeq 1~\rm cm^{-3}$; 2) transport of CRs in the Galaxy is mainly diffusive, with a momentum dependent diffusion coefficient $D(p)$ over a magnetized halo of size $H$, where the gas density is negligible. In this simple picture the propagation time is dominated by diffusion in the halo, $\tau_{d}(p)=H^{2}/D(p)$, but most of the grammage is accumulated by CRs when they traverse the disc. Hence the grammage can be written as $X(p)\approx 1.4 m_{p} n_{d} (h/H) v(p) \tau_{d}$, where $m_{p}$ is the proton mass, $v(p)\sim c$ is the particle velocity and the numerical factor accounts for the fact that both protons and helium nuclei in the interstellar medium serve as target for the particle collisions that define the grammage $X(p)$. The quantity $n_{d}(h/H)$ is the mean density experienced by CRs while diffusing in an almost empty halo and occasionally traversing the dense disc. The best measurement of the grammage comes from the ratio of the fluxes of Boron and Carbon nuclei (B/C): Boron is mainly a secondary product of the nuclear collisions of Carbon (and heavier) nuclei, while C is instead mainly primary, namely directly accelerated in the sources of CRs. The B/C ratio is proportional to the grammage and scales as $1/D(p)$, where it is assumed that the diffusion coefficient is the same throughout the disc and halo regions. Many refinements of this simplest picture are available in the literature, but the general scenario stays qualitatively as outlined above.

This simple diffusive description of the transport is expected to become increasingly more accurate when applied to spatial scales much larger than the coherence scale of the Galactic magnetic field, usually taken to be in the $50-100$ pc range \cite{haverkorn}. In contrast, this picture is known to be inadequate to describe the transport of CRs on smaller scales, as it is for instance the case for the central pc scale region around the Galactic center or the close surroundings of a source. In other words, the diffusion paradigm requires averages on several correlation lengths of the turbulent magnetic field responsible for CR diffusion. In the region immediately outside a CR source the propagation of CRs may lead to a very elongated shape of the cloud of particles diffusing away from the source \cite{nava,giacinti}, reflecting the quasi-ordered structure of the Galactic magnetic field on a scale $L_c$. However, in these investigations, CRs were assumed to be passive actors diffusing in a pre-established turbulent field. Some investigations of the non-linear problem in which CRs affect their own diffusion have been presented in Refs.~\cite{plesser,malkov}. While the main focus of Ref.~\cite{malkov} was on describing the possible interaction of escaping particles with a nearby molecular cloud magnetically connected with the CR source, a SNR,  in Ref.~\cite{plesser} a problem very similar to the one studied in this article was considered, though with different boundary conditions and more restrictive assumptions. Scenarios involving non-linear CR diffusion on Galactic scales have also been recently discussed \cite{blasiPRL,aloJCAP}. At the same time, phenomenological propagation models assuming a diffusion coefficient that close to the sources is different from the Galactic average have been sometimes explored in the literature (see {\it e.g.} \cite{cowsik} and references therein) as a possible explanation of the observed energy dependence of the flux of secondaries.

In this paper we study the CR propagation close to their sources, with a special attention for the implications of this process on our understanding of global CR propagation on Galactic scales. The density of CRs leaving their parent SNR stays larger than the Galactic mean density for relatively long times, even in the absence of non-linear effects. The strong spatial gradient in the CR density generates hydromagnetic waves (streaming instability), that in turn slow down the CR propagation in the near-source region. Within a distance of about 1-2~$L_c$, with $L_c=50-150$ pc\cite{haverkorn}, the problem can be well approximated as one dimensional (see also \cite{nava,giacinti,malkov}). We show that for particle energies up to a few TeV, the grammage accumulated by CRs within such distances, as due to non-linear diffusive transport in the dense Galactic disc ($n_{d}\sim 1~\rm cm^{-3}$), may become comparable with the global grammage expected in the standard picture of propagation throughout the whole Galaxy, as deduced from the measurement of the B/C ratio. The implications of this finding for our understanding of the origin of CRs will be discussed. 

\vskip .3cm
{\it Calculations} -- As a benchmark for the Galactic diffusion coefficient we adopt the functional form $D_{g}(E)=3.6\times 10^{28} E_{GeV}^{1/3}~\rm cm^{2}/s$, as derived in Ref.~\cite{fit} from a leaky-box fit to GALPROP  \cite{galprop} (see http://galprop.stanford.edu) results for a Kolmogorov turbulence spectrum (here, for simplicity, we restrict ourselves to the relativistic regime). The scenario we have in mind is that of a supernova (SN) that explodes in the Galactic disc, where the magnetic field is assumed to have a well established direction over the coherence length $L_{c}\sim 50-150$ pc. In fact the magnetic field direction will not experience dramatic changes even on scales somewhat larger than $L_{c}$ if the turbulence level is low, $\delta B/B<1$. Describing the particle transport as diffusive on scales $\lesssim L_{c}$ can only be done for particles with a mean free path $3D_g(p)/c<<L_c$. This condition is easily seen to be satisfied up to at least $\sim 10^{5}-10^{6}$ GeV for the standard Galactic diffusion coefficient $D_{g}$: we will only be concerned with particles well below this energy. After a time $\sim L_{c}^{2}/D_{g}(E)\sim 9\times 10^{4} E_{GeV}^{-1/3}$ years, particles start diffusing out of the region where the magnetic field can be assumed to have a given orientation and the problem should be treated as 3-dimensional diffusion. In such a phase, within a distance from the source $\sqrt{D_g(E)t_s}$, the CR density due to the source itself remains larger than the mean galactic density for a time $t_{s}$ that we can estimate by equating the individual source contribution, $N(E)/(4\pi D_{g}(E) t)^{3/2}$, to the average Galactic density, $N(E){\cal R}H/(2\pi R_{d}^{2 }D_{g}(E))$, with $N(E)$ the average spectrum that a source of CRs injects in the Galaxy, ${\cal R}$ the SN rate, $R_d$ and $H$ the size of the galactic disc and halo respectively. For typical values of the parameters, ${\cal R}=1/30~\rm yr^{-1}$, $R_{d}=30$ kpc (from \cite{fit}) and $H=4$ kpc, one finds $t_s\sim 2\times 10^{4} E_{GeV}^{-1/3}\ yr$, which indicates that the density of locally accelerated CRs quickly drops to the galactic average as soon as propagation becomes 3-dimensional.
Hence we formulate our problem starting from the solution of the one-dimensional transport equation,
\be
\frac{\partial f}{\partial t} + v_{A} \frac{\partial f}{\partial z}- \frac{\partial}{\partial z} \left[ D(p,z,t) \frac{\partial f}{\partial z}\right] = q_{0}(p)\delta(z)\Theta(T_{SN}-t)\ ,
\label{eq:transport}
\ee
in a box of size $2L_{c}$ with the boundary condition that $f(p,|z|=L_{c},t)=f_{g}(p)$, with $f_g$ the diffuse CR spectrum in the Galaxy. Eq.~\ref{eq:transport} is meant to describe a situation in which diffusion is due to self-generated waves moving away from the source at the Alfv\'en speed $v_{A}$ (second term on the {\it lhs} of the equation).
 
Injection is assumed to be constant in time from $t=0$ to a time $T_{SN}$, which characterizes the duration of the release phase of CRs into the ISM. Since we are interested in CRs with energies below $\sim 100$ TeV or so (for higher energies the density of particles close to the source is too small to lead to effective growth of the streaming instability), the escape of CRs is expected to occur at the time of shock dissipation. The function $q_{0}(p)=A\left(p/ m_{p}c\right)^{-4}$ mimics injection at a strong SNR shock, with the normalisation constant $A=\xi_{CR} E_{SN}/\pi R_{SN}^{2} T_{SN} {\cal I}$, and ${\cal I} = \int_{0}^{\infty} dp 4\pi p^{2}\left( p/m_{p}c\right)^{-4}\epsilon(p)$, where $\epsilon(p)$ is the kinetic energy of a particle with momentum $p$. The normalization is such that a fraction $\xi_{CR}$ of the kinetic energy $E_{SN}$ of the SNR shock is converted into CRs. The radius of the SNR at the time of escape of CRs is chosen to be $R_{SN}\approx 20$ pc, of order the size of the slowly varying radius of a SNR during the Sedov phase in the ISM. Integrating Eq.~ \ref{eq:transport} in a neighborhood of $z=0$ one finds:
\be
\left .\ \frac{\partial f}{\partial z}   \right  |_{z=0}= - \frac{q_{0}(p)\Theta(T_{SN}-t)}{2 D(p,z,t)|_{z=0}},
\ee
which is used as a boundary condition on Eq. \ref{eq:transport}.

The diffusion coefficient in Eq. \ref{eq:transport} is self-generated by CRs leaving the source:
\be
D(p,z,t) = \left .\ \frac{1}{3} r_{L}(p) v(p) \frac{1}{\cF (k,z,t)} \right |_{k=1/r_{L}(p)} ,
\label{eq:diff}
\ee
where the spectrum of the self-generated waves $\cF (k,z,t)$ satisfies the differential equation:
\be
\frac{\partial \cF}{\partial t} + v_{A}\frac{\partial \cF}{\partial z} = (\Gamma_{CR} - \Gamma_{D}) \cF\ .
\label{eq:waves}
\ee
In the latter equation,
\be
\Gamma_{\rm CR}(k)=\frac{16 \pi^{2}}{3} \frac{v_{\rm A}}{\cF B_{0}^{2}} \left[ p^{4} v(p) \frac{\partial f}{\partial z}\right]_{p=q B_{0}/kc}\,
\label{eq:gammacr}
\ee 
is the growth rate of the resonant streaming instability associated with CRs moving at superalfv\`enic speed \cite{skilling}, while the damping rate $\Gamma_D=\Gamma_{IN}+\Gamma_{NLD}$ contains both the effects of ion-neutral damping (IND) at rate $\Gamma_{IN}$ \cite{kulsrud} and non-linear wave damping (NLD) \cite{landau}. For the rate of NLD we use:
\be
\Gamma_{NLD} = (2 c_{K})^{-3/2} k v_{A} \cF^{1/2}~~~~~c_{K}\approx 3.6\ ,
\label{eq:damp}
\ee
which is equivalent to assuming a Kolmogorov description of the cascade. Another source of damping, suggested in Ref.~\cite{FG04}, is provided by the pre-existing MHD turbulence, which these authors assume to cascade anisotropically in $k$ from a large injection scale $L_{\rm MHD}$, comparable with our $L_c$. This contribution is included in our calculations but it has little effect on the global solution.

The relative importance of IND and NLD depends on the abundance of neutral atoms in the region surrounding the SN. The issue of the role of IND for CR propagation in the Galaxy is all but new: \cite{skilling,holmes} pointed out that IND would induce a wave free zone above and below the Galactic disc where CRs would move ballistically. Diffusion would be granted in a far region, where wave generation would be faster than IND and particle isotropization would take place in such region. The importance of IND for particle scattering was also discussed in Ref. \cite{kulsrud}. After this pioneering work, the role of IND for CR propagation was almost completely ignored and at present most calculations of CR transport assume that diffusion is present everywhere in the halo and disc and its origin is not much disputed. One of the reasons of this attitude towards the problem of IND is that the neutral material may be speculated to be spatially segregated in dense regions with small filling factor, while most of the propagation volume would be filled with rarefied ionized gas. Following \cite{ferriere}, one could argue that the part of ISM that is most important in terms of particle propagation is a mix of the warm ionized and hot ionized medium (WIM and HIM), while IND would severely inhibit wave growth in the rest of the ISM, thereby suppressing diffusion. On the other hand, following \cite{jean}, one could argue that in the WIM most gas is ionized with density that can be as high as $\sim 0.45 \rm cm^{-3}$ but neutral gas is still present with density $\sim 0.05 \rm cm^{-3}$. Such parameters would still be more than enough to quench the growth of Alfv\'en waves due to IND on time scales of relevance for propagation of CRs in the Galaxy. In the region around a SNR, the level of IND inferred for these parameters is still important but not as much as for the Galactic CR propagation (see below). Moreover, as argued in Ref. \cite{ferriere}, the WIM, having temperature $\sim 8000$ K, is expected to be made of fully ionized hydrogen, while only helium would be partially ionized. This latter picture would have prominent consequences in terms of IND, in that this process is due to charge exchange between ions and the partially ionized (or neutral) component, but the cross section for charge exchange between H and He is about three orders of magnitude smaller \cite{aladdin} than for neutral and ionized H, so that the corresponding damping rate would be greatly diminished. Unfortunately, at present, there is no quantitative assessment of this phenomenon and we can only rely on a comparison between cross sections of charge exchange. On the other hand, it is also possible that a small fraction of neutral hydrogen is still present, in addition to neutral helium: following \cite{ferriere}, the density of neutral H is $\lesssim 6\times 10^{-2} n_{i}$, and for $n_{i}=0.45  \rm cm^{-3}$ this implies an upper limit to the neutral H density of $\sim 0.03 \rm cm^{-3}$. 

In order to account, to some extent, for the uncertainty in the role of IND around sources of CRs, below we consider the following cases: (1) No neutrals and gas density $0.45~\rm cm^{-3}$ (this is the best case scenario in terms of importance of the near source grammage, with the underlying assumption that the neutral component is entirely in the form of He); (2) Neutral density $n_{n}=0.05~\rm cm^{-3}$ and ion density $n_{i}=0.45~\rm cm^{-3}$; (3) Density of ionized H $0.45~\rm cm^{-3}$ and density of neutral H $0.03~\rm cm^{-3}$; (4) rarefied totally ionized medium with density $n_{i}=0.01~\rm cm^{-3}$. 

In order to avoid artificial divergences in the diffusion coefficient we assume that there is always a minimum background of waves $\cF_0$ that corresponds, through Eq.~\ref{eq:diff}, to the mean Galactic diffusion coefficient $D_{g}$. For most locations and times around the source, self-generated waves exceed the pre-existing ones up to scales corresponding to particle energies $\lesssim 10$ TeV, above which the streaming instability grows too slowly and the diffusion coefficient is well described by $D_{g}$, as we show below. 

Eqs.~\ref{eq:transport} and \ref{eq:waves} are solved together using a finite differences scheme with backward integration in time. Given the non-linear nature of the problem, the diffusion coefficient in the second term of Eq. \ref{eq:transport} is evaluated at time $t$ while the spatial derivatives of the distribution function are evaluated at time $t+\delta t$, where $\delta t$ is the time step used for numerical integration. 

Due to the non-linear nature of the problem the calculation of the grammage traversed by the particles is not straightforward. We devised the following procedure that provides an estimate of the average particle residence time based on measurement of the particle flux across the boundaries at distance $L_c$ from the source. In the case of propagation in the preassigned diffusion coefficient $D_{g}$, we find that injection of a burst of particles at $t=0$ leads to $\approx89\%$ of them leaving the system through the boundaries at $\pm L_c$ within a time $L_{c}^{2}/D_{g}$ (the classical estimate of the diffusion time for such a configuration). In the non-linear case, the flux of particles across the boundaries, integrated over a time $t$, $\phi(t) = -2 \int_{0}^{t} dt D \frac{\partial f}{\partial z}|_{z=L_{c}}$, is calculated numerically. The average time $T_{d}$ by which $\sim 89\%$ of the particles leave the region of size $2L_{c}$ around the source is estimated by requiring that $\phi(T_{d}) = 0.89 q_{0} \rm Min\left[T_{d},T_{SN}\right]$, in analogy with the case of preassigned diffusion coefficient. The estimate eventually fails when the propagation time becomes comparable with the duration $T_{SN}$ of the injection phase. In the non-linear regime, this typically happens at energies above $\gtrsim 10$ TeV. The grammage traversed by particles while leaving this near-source region can be written as $X_{s}(p)\approx 1.4 m_{p} n_{d} T_{d} v(p)$, to be compared with the grammage usually associated with propagation in the Galaxy.

\begin{figure}
\begin{center}
{\includegraphics[width=\linewidth]{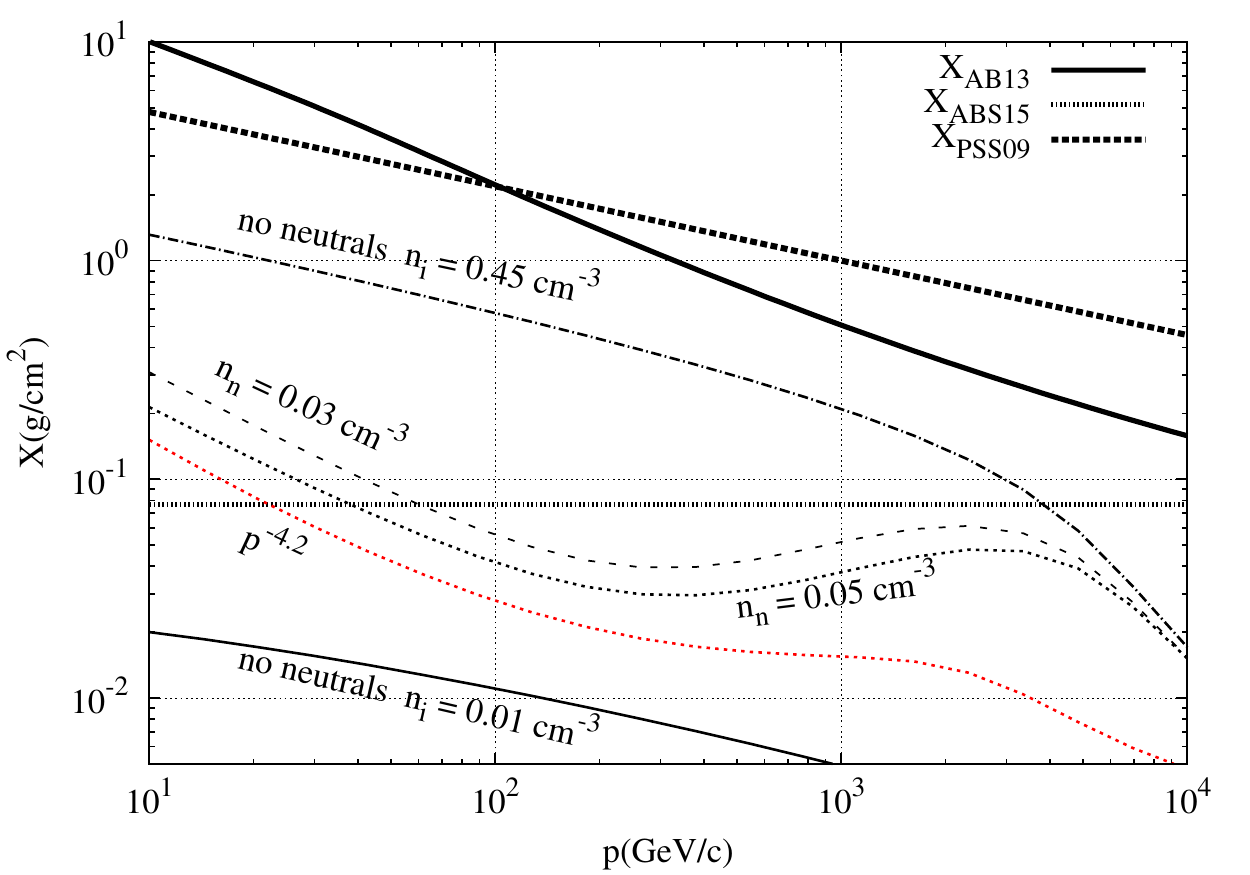}}
\caption{Grammage accumulated by CRs in the near-source region for $L_{c}=100$ pc in the three cases: (1) $n_{n}=0$, $n_i = 0.45 \mathrm{cm^{-3}}$; (2) $n_i = 0.45 \mathrm{cm^{-3}}$ and $n_n = 0.05 \mathrm{cm^{-3}}$; (3) $n_{n}=0$, $n_i = 0.01 \mathrm{cm^{-3}}$, as labelled. The thin dotted (red) line corresponds to case (2) but with slope of the injection spectrum 4.2. The thick dashed line (labelled as $X_{\rm PSS09}$) shows the grammage inferred from the measured B/C ratio \cite{fit}, while the thick solid line (labelled as $X_{\rm AB13}$) shows the results of the non-linear propagation of Ref. \cite{aloJCAP}. The horizontal (thick dotted) line (labelled as $X_{\rm ABS15}$) is the source grammage, as estimated in Ref. \cite{serpico}.}
\label{fig:grammage}
\end{center}
\end{figure}

\vskip .3cm
{\it Results} --  The grammage $X$ as a function of particle energy (or rigidity) is plotted in Fig.~\ref{fig:grammage}, for the case of a SNR with total kinetic energy $E_{SN}=10^{51}$ erg. The field coherence length is taken to be $L_c=100$ pc and the CR acceleration efficiency is $\xi_{CR}=20\%$. In terms of properties of the medium around the source, the four cases mentioned above are considered (see labels on the curves). 
The thick dashed line shows the grammage estimated from the measured B/C ratio, assuming standard CR propagation in the Galaxy with turbulence described {\it a la} Kolmogorov \cite{fit}. The thick solid curve represents the grammage as calculated in the model of non-linear CR propagation of Ref.~\cite{aloJCAP} (see also \cite{blasiPRL}), while the horizontal (thick dotted) line shows an estimate of the grammage traversed by CRs while still confined in the downstream of the SNR shock \cite{serpico}. In all cases we assumed injection $\propto p^{-4}$, but for case (2) above we also considered the case of steeper injection (thin dotted (red) line).

In case (1), in the energy region $E\lesssim$ few TeV, the grammage contributed by the near-source region due to non-linear effects is comparable (within a factor of $\sim$ few) with that accumulated throughout the Galaxy if the standard diffusion coefficient is adopted. When neutral atoms are present, the IND severely limits the waves' growth: in case (2) the grammage in the energy range $E\lesssim$ few hundred GeV is about ten times smaller. However, since the importance of IND decreases with decreasing wavenumber $k$, particles at energies above $\sim 1$ TeV are again allowed to generate their own waves and the grammage in the near-source region increases, thereby becoming comparable with the one accumulated inside the source.

As pointed out above, following \cite{ferriere}, it seems plausible that most of the neutral gas in the warm-hot phase is made of helium, whose charge exchange cross section with ionized hydrogen is very small. Ref. \cite{ferriere} suggests an upper bound to the density of neutral hydrogen of $\sim 0.03 \rm cm^{-3}$. This case is accounted for as Case (3) above.

Case (4) corresponds to a small grammage (due to the low gas density) but it is important to realize that in fact Cases (1) and (4) correspond to roughly the same propagation time in the near-source region. This might have important observational consequences in the case in which a dense target for $pp$ collisions, such as a molecular cloud, is present in a region where the gas density (outside the cloud) is very low and IND is absent: the long escape times and the correspondingly enhanced CR density will reflect in enhanced gamma-ray emission.

\begin{figure}
\begin{center}
{\includegraphics[width=\linewidth]{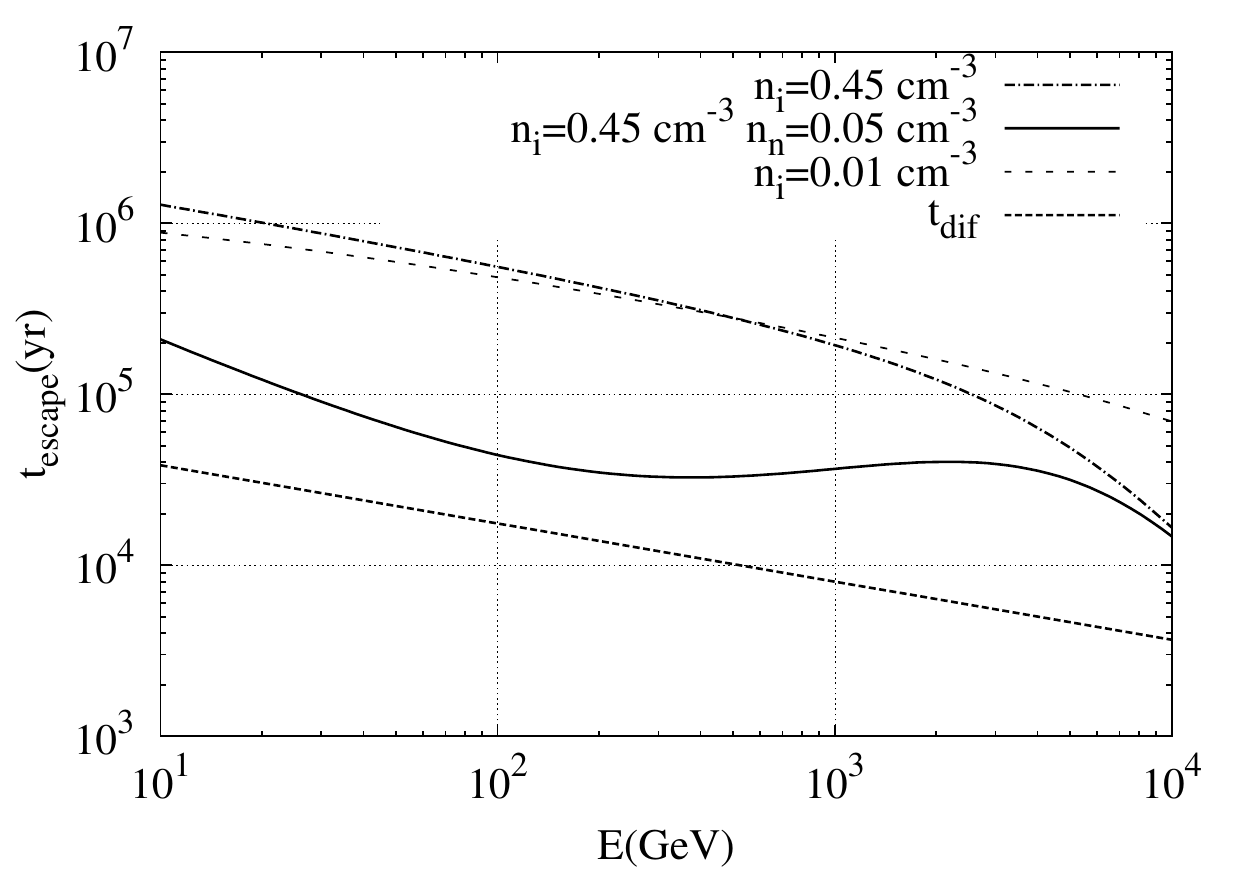}}
\caption{Escape time of CRs from the near-source region for $L_{c}=100$ pc in the three cases: (1) $n_{n}=0$, $n_i = 0.45 \mathrm{cm^{-3}}$; (2) $n_i = 0.45 \mathrm{cm^{-3}}$ and $n_n = 0.05 \mathrm{cm^{-3}}$; (3) $n_{n}=0$, $n_i = 0.01 \mathrm{cm^{-3}}$. The dotted line refers to the escape time calculated using the Galactic diffusion coefficient $D_{g}$.}
\label{fig:Tescape}
\end{center}
\end{figure}

The time needed for CR escape from the region of size $L_{c}=100$ pc around a source is shown in Fig.~\ref{fig:Tescape} for the four cases of interest, compared with the diffusion time in the same region estimated by using the Galactic diffusion coefficient $D_{g}$ (dotted line). This plot shows once more that the escape time is weekly dependent upon the density of ions provided there is no appreciable IND. The small difference between the two cases (dash-dotted and dashed lines) is to be attributed to the weak advection with Alfv\'en waves, since the waves' velocity is somewhat different in the two cases.

In the absence of neutrals, the near-source grammage increases with increasing $L_{c}$ and with increasing CR acceleration efficiency $\xi_{CR}$, proportional to $\sim L_{c}^{2/3}$ and $\propto \xi_{CR}^{2/3}$ respectively. It is interesting to notice that these trends are the same shown by the self-similar solution obtained in Ref.~\cite{plesser} for a similar problem, though with different boundary conditions and under the assumption of impulsive CR release by the source. In the cases in which neutral atoms are absent, for particles with energies up to $\sim 1$ TeV, the grammage decreases with energy in roughly the same way as the observed grammage \cite{fit}, as a result of the dependence of the NLD rate on $k$ in Eq. \ref{eq:damp}.

\begin{figure}
\begin{center}
{\includegraphics[angle=0,width=0.9\linewidth]{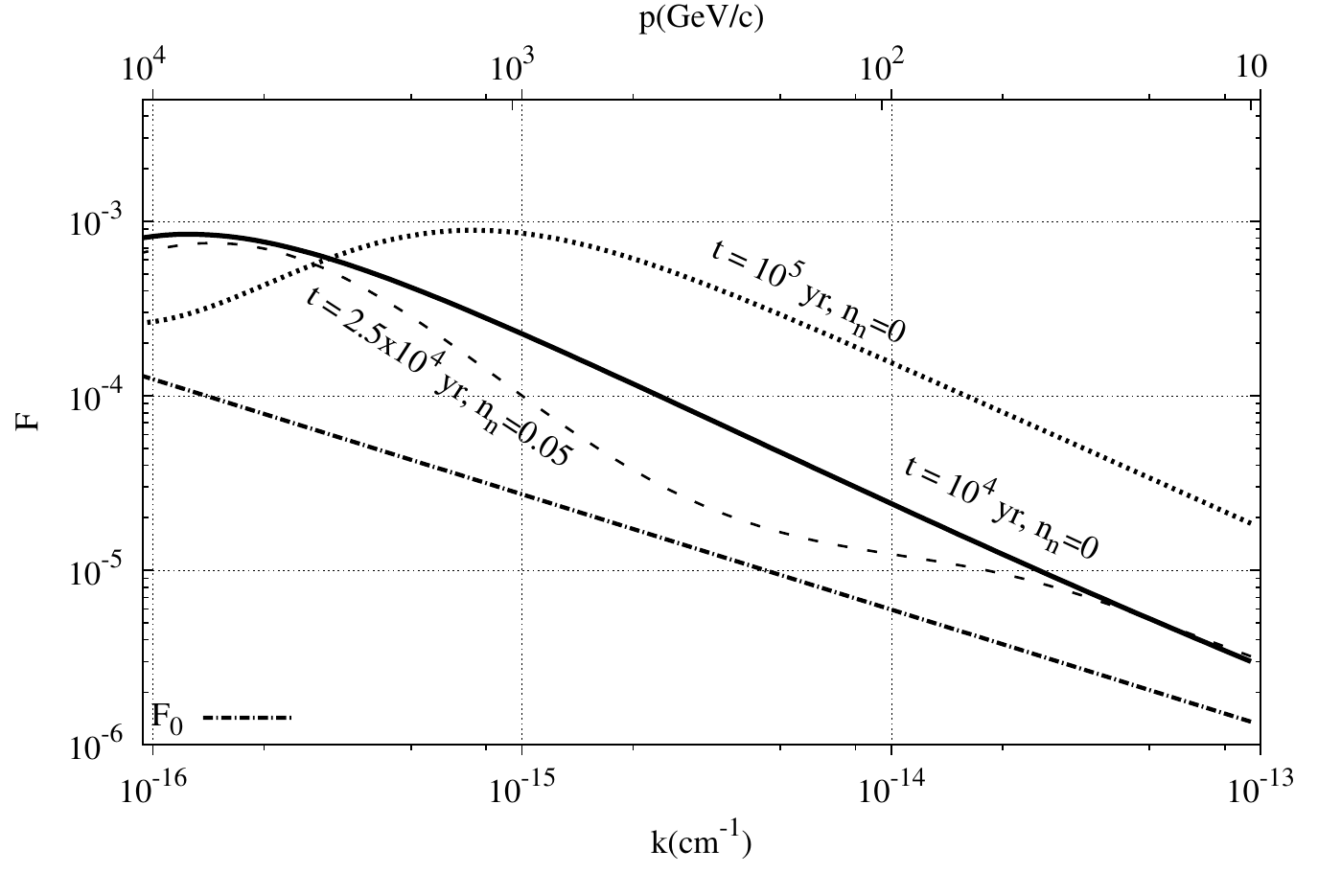}}
\caption{Normalized power spectrum of waves for $L_{c}=100$ pc, at distance $z=50$ pc from the source, $10^{4}$ and $10^{5}$ years after CR release in the ISM. The solid and dotted lines correspond to the fully ionized medium with $n_i = 0.45 \mathrm{cm^{-3}}$, while the dashed line corresponds to the warm ionized medium with $n_i=0.45 \mathrm{cm^{-3}}$ and $n_n=0.05 \mathrm{cm^{-3}}$. The dot-dashed line shows the Galactic power spectrum, corresponding to the diffusion coefficient that provides a good fit to the GALPROP grammage \cite{fit}.}
\label{fig:waves}
\end{center}
\end{figure}

The enhanced grammage illustrated in Fig.~\ref{fig:grammage} is the result of streaming instability excited by CRs leaving the source. This effect is particularly important for particles with energy $\lesssim 10$ TeV, because of the large density of particles at such energies, that reflects into a correspondingly high growth rate of the instability (see Eq.~\ref{eq:waves}). In Fig.~\ref{fig:waves} we show the power spectrum $\cF (k)$ at $z=50$ pc for a case with $L_{c}=100$ pc. On the top x-axis we show the momentum of particles that can resonate with waves of given wavenumber $k$ (bottom x-axis). The solid (dashed) line refers to case (1) at time $t=10^{4}$ ($t=10^{5}$) years. In Case (2), the presence of neutrals decreases the level of self-generated waves (see dotted line, computed at $t=10^{4}$ years), which however remains appreciably higher than the Galactic turbulence level $\cF_{0}(k)$, also shown in Fig.~\ref{fig:waves} as a thick dot-dashed curve.
Particles diffusing away from the source keep pumping waves into the environment for about $10^{5}$ years. At later times, higher energy particles start escaping the near-source region, the gradients diminish and $\cF(k)$ approaches again $\cF_{0}(k)$, starting from low values of $k$.

\begin{figure}
\begin{center}
{\includegraphics[angle=0,width=0.9\linewidth]{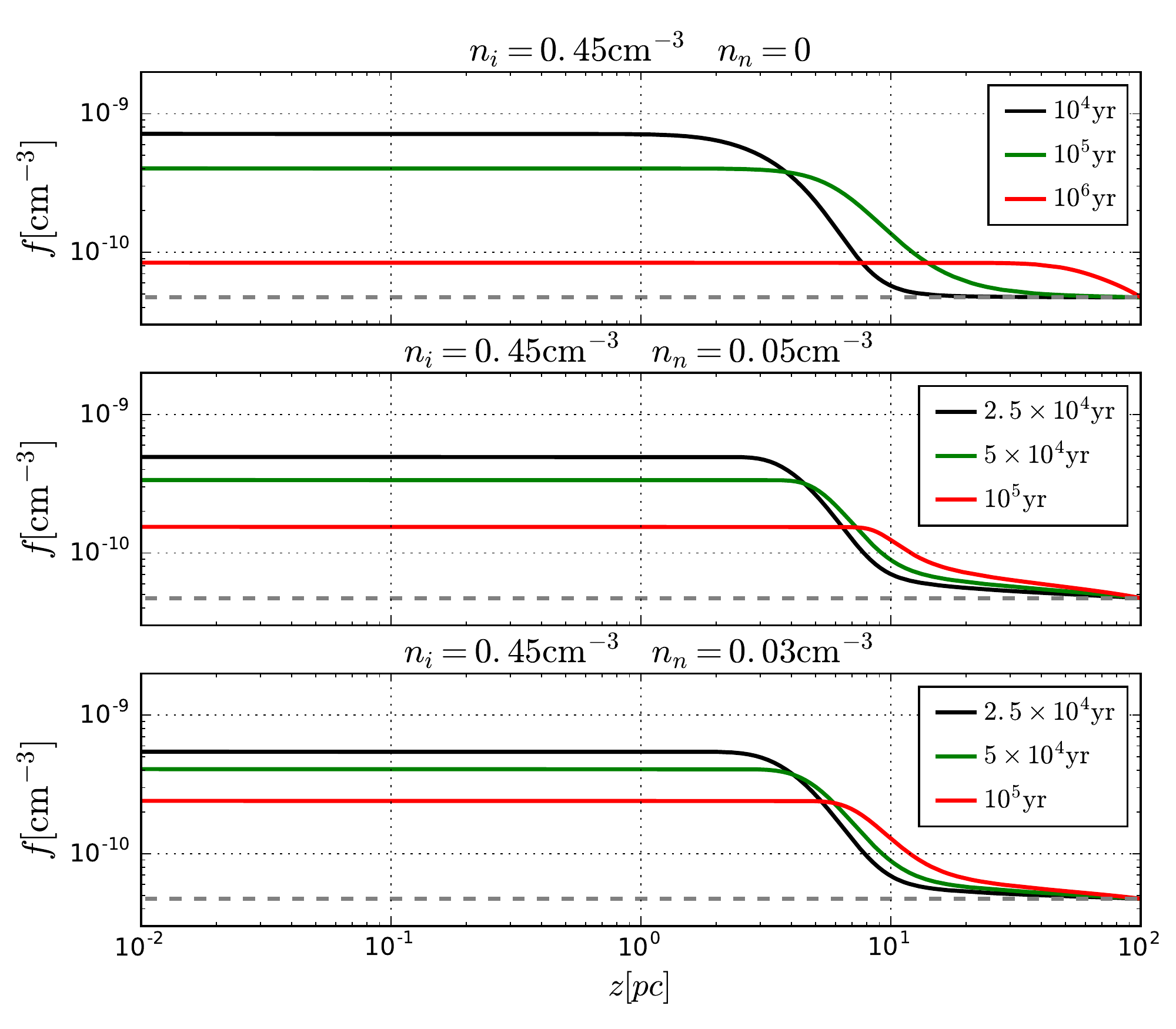}}
\caption{Particle density $4\pi p^{3} f(p)$ for $p=10$ GeV/c as a function of distance from the source. The curves refer to times $10^{4}$, $10^{5}$ and $10^{6}$ years, as labelled (we used $L_{c}=100$ pc). {\it Top Panel:} $n_{i}=0.45\rm cm^{-3}$ and $n_{n}=0$. {\it Bottom Panel:} $n_{i}=0.45\rm cm^{-3}$ and $n_{n}=0.05\rm cm^{-3}$.}
\label{fig:density}
\end{center}
\end{figure}

The effect of particle self-confinement is illustrated in Fig.~\ref{fig:density}, where we show the density of particles  (or more correctly the quantity $4\pi p^{3} f(p)$) with momentum $p=10$ GeV/c as a function of the distance from the escape surface, for three times after release ($10^{4}$, $10^{5}$ and $10^{6}$ years). The top and bottom panels refer to the cases with no neutrals and ion density $n_{i}=0.45\rm cm^{-3}$, and ion density $n_{i}=0.45\rm cm^{-3}$ with neutral density $n_{n}=0.05\rm cm^{-3}$ respectively. In the latter case the effect of IND is that of limiting wave growth and reducing the time needed for particle escape from the near-source region. One can clearly see that on time scales of order $10^{4}$ years, comparable with the diffusion time on a scale $L_{c}=100$ pc with Galactic diffusion coefficient $D_{g}$, the density of particles in the near-source region ($\sim 1-2$ pc) remains more than one order of magnitude larger than the Galactic background in the case with no neutrals, and about one order of magnitude larger in the case with $n_n=0.05 cm^{-3}$. Even after $10^{6}$ years the density of particles in the near-source region ($\sim 30-50$ pc) remains appreciably larger than the Galactic mean density in the case of no neutrals (or neutrals consisting of He alone).

Eventually all particles injected by the source escape the region, the density of CRs drops to the mean Galactic value (assumed to be the one observed at the Earth) and the wave power spectrum drops to $\cF_{0}(k)$ at all values of $k$. One should notice that even if the effect of CR streaming is that of increasing the energy density of waves by orders of magnitude, typically it remains true that $\cF(k)\lesssim 1$, hence the use of quasi-linear theory remains well justified.
\vskip .3cm
{\it Discussion} -- We calculated numerically the grammage traversed by CRs while propagating in the disc of the Galaxy in the region immediately outside the source, assumed to be a typical SNR with a total energy of $10^{51}$ erg and a CR acceleration efficiency of $20\%$. On scales of order the coherence scale of the Galactic magnetic field, $L_{c}\sim 100$ pc, the problem can be considered as one-dimensional. At larger distances diffusion becomes 3-dimensional and the density of CRs contributed by the individual source quickly drops below the Galactic average (assumed to be the one observed at the Earth). The gradient in the spatial distribution of CRs in the near-source region is responsible for generation of Alfv\'en waves that in turn scatter the particles, thereby increasing their residence time in the near-source region. This phenomenon was previously studied in Ref.~\cite{malkov} in connection with the problem of the interaction of CRs with molecular clouds in the near source region, and in Ref.~\cite{plesser}, who studied the consequences of impulsive release of CRs by a source embedded in a fully ionised medium.

We showed that in this scenario, the grammage traversed by CRs with energies up to a few TeV is heavily affected by the self-induced confinement close to the sources, to an extent that depends on the number density of neutral hydrogen in the Galactic disc. In the absence of neutrals and with density of $\sim 0.45~\rm cm^{-3}$, the near-source grammage is comparable (within a factor of few) with the one that is usually inferred from the measured B/C ratio at the position of the Earth. Clearly this does not imply that the propagation time of CRs in the near-source region is close to the escape time of CRs from the Galaxy, because of the higher gas density in the disc, a factor $H/h$ times larger than the average density traversed by CRs while diffusing in the disc+halo region. The fact that the residence time in the near source region remains relatively short compared with the overall residence time in the Galaxy, in the considered energy range, leads us to expect that the effects on radioactive isotopes (the other observable that is used to infer the propagation time of CRs in the Galaxy in addition to the secondary/primary ratios) will be negligible. When neutrals are present, as expected for the standard interstellar medium, the near-source grammage is lower, but the strength of the damping is not well established, mainly because in the warm-hot interstellar medium most of the neutral gas is expected to be in the form of He atoms. The cross section for H-He charge exchange is much smaller than for H-H, therefore it seems plausible to assume that most damping is actually due to a residual fraction of neutral H that may be left. Here we made an attempt to bracket the importance of IND as due to all these uncertainties.

At energies $\gtrsim 1$ TeV the grammage traversed by CRs can reasonably be expected to be heavily affected by the in-source contribution, due to the fact that CRs are trapped in the downstream region of the SN shock before escaping, as already proposed in Ref. \cite{serpico} (where the confinement time is assumed to be $\sim 10^{4}$ years). This is probably not the case at very high energies, around the knee, since such CRs are expected to escape the source from the upstream region at earlier times \cite{dam}.

There are at least two possible signatures which relate CR residence time in the near-source region and experimental data: the formation of extended halos of gamma ray emission from $\pi^0$ decay in a region of size $\sim L_c$ around CR sources and enhanced inverse Compton scattering and Synchrotron emission from electrons. Both these aspects will be discussed in a forthcoming paper.
The general picture that arises from these considerations is that at all energies the observed grammage is affected by either non-linear propagation in the near-source region or transport inside the source, thereby making the translation of the grammage (from B/C) to a confinement time in the Galaxy rather problematic. More specifically, both the normalisation and the slope of the inferred diffusion coefficient are likely to reflect more a combination of Galactic propagation plus one of the two phenomena described above rather than the pure CR transport on Galactic scales. 

{\it Note}: During the review process of the present paper, a different group published the results of numerical propagation of CRs in the near source region in the presence of non-linear wave generation, reaching qualitatively similar conclusions \cite{nava1}. Their work focuses more on the implications of non-linear CR transport for gamma ray observations in the occasional presence of a molecular cloud near the source, while we investigate the implications for CR grammage, and include the case of negligible IND, as could be expected in a medium where all H is fully ionized and the neutral component is only made of He.

\vskip .3cm
{\it Acknowledgments} -- This work was partially funded through grant PRIN-INAF 2013.

\end{document}